\documentclass[twocolumn,showpacs,amsmath,amssymb,prx,graphics,graphicx
]{revtex4-1}
\usepackage{bm}
\usepackage{graphicx}
\usepackage{dcolumn}
\begin{document}
%\date{Dec.1999 - Aug.2000 - \today}
\title
{Determining the vortex tilt relative to a superconductor surface} 
%crossing the $ab$ surface of a uniaxial  superconducting half-space}
\author{V. G. Kogan}
\affiliation{Ames Laboratory DOE and Iowa State University, Ames, IA 50011, USA }
\author{J. R. Kirtley}
\affiliation{Geballe Laboratory for Advanced Materials, Stanford University, Stanford, CA 94305, USA}
\begin{abstract}
 It is of interest to determine the exit angle of a vortex from a superconducting surface, since this affects the intervortex interactions and their consequences. 
 %, for example, can be a determining factor in its pinning strength. 
 Two ways to determine this angle are to image the vortex magnetic fields above the surface, or the vortex core shape at the surface. In this work we evaluate the field $\bm h(x,y,z)$ above a flat superconducting surface  $x,y$ and  the currents $\bm J(x,y)$ at that surface for a straight vortex tilted relative to the normal to the surface, for both the isotropic and anisotropic cases.  In principle, these results can be used to determine the vortex exit tilt angle from analyses of magnetic field imaging or density of states data.    \end{abstract}
%\pacs{74.20.-z,74.78.-w,74.78.Fk,74.81.-g}
\date\today
\maketitle
 
 \section{Introduction}
 
%Among various methods of studying vortices and vortex lattices the surface probes provide direct local information on the field distribution (Scanning SQUID Microscopy) and the density of states (Scanning Tunneling Microscopy).  
 
 In the long history of studying vortices and vortex lattices  with the help of  surface probes (e.g. Bitter decoration \cite{Essmann,dolan1989prl}, Hall bar microscopy \cite{Chang1992apl}, magnetic force microscopy \cite{Hug1994} scanning Superconducting QUantum Interference Device microscopy (SSM) \cite{kirtley2010rpp}, or scanning tunneling microscopy (STM) \cite{Hess}) vortices were commonly assumed to exit the superconductor perpendicular to the surface. H. Hess and collaborators were the first to examine vortex lattices in NbSe$_2$ in tilted fields \cite{Hess} using STM. They found a peculiar ``comet-like" density of states (DOS) distribution near the vortex core. Recently, the STM group of H. Suderow concluded that the vortex lattice structure in  fields tilted relative to a plane surface of nearly isotropic $\beta$-Bi$_2$Pd is affected by the surface contribution to the vortex-vortex interactions  due to vortex stray fields outside the sample \cite{Jesus}.
 
The question then arises as to whether one can determine the vortex orientation relative to the surface by measuring the field above the sample surface or the DOS at the surface for a superconductor containing vortices. This question is addressed in this paper.

 A    uniaxial crystal with a surface in an arbitrary crystal plane and a vortex oriented arbitrarily relative to the crystal have been considered in \cite{KSL} within the general anisotropic London approach. The formalism ``generality" in this paper made the outcome quite cumbersome 
  and not easily applied. Besides, it was unclear what features of the field distribution outside, or of the DOS at the interface, are due to the vortex tilt and which are due to crystal anisotropy. 
 
 For this reason, we focus here first on the isotropic half-space superconductor at $z<0$ and a straight vortex approaching the interface $z=0$ at an angle $\theta$ with the normal $\hat z$ to the surface. For $\theta=0$ this problem has been solved by J. Pearl \cite{Pearl}. We find that even in the isotropic case, the field distribution above the surface and the currents $\bm J(x,y)$ flowing at the surface carry measurable signs of the vortex tilt. The stray field $h_z(x,y;z)$ can, in principle, be measured by field sensitive probes such as scanning Hall bar or scanning SQUID, whereas $|J(x,y)|$ affects the pair potential and DOS probed by STM. 
 
 In the second part of this paper, we consider tilted vortices in   uniaxial superconductors with the surface in the $ab$ plane.  
 
 \section{Isotropic case}
 
 The field $\bm h$ outside the superconductor satisfies div$\,\bm h$ = curl$\,\bm h$ = 0, so that one can look for this field as $\bm h=\bm\nabla \varphi$ with the potential $\varphi$ obeying the Laplace equation $\Delta\varphi=0$. The general solution of this equation in the upper half-space $z>0$  for $\varphi\to 0$ as $z\to\infty$ is:
 \begin{eqnarray}
 \varphi (\bm r,z)= \int \frac{d^2\bm k}{4\pi^2}\,\varphi(\bm k )\,e^{i\bm k\bm r-kz} \,, \label{phi}
\end{eqnarray}
 where 
  \begin{eqnarray}
 \varphi (\bm k)e^{ -kz}= \int  d^2\bm r \,\varphi(\bm r,z )\,e^{-i\bm k\bm r }  \label{phi_k}
\end{eqnarray}
is the two dimensional (2D) Fourier transform of the potential $\varphi$.
 
Inside the superconductor, the field components satisfy  the London equations
  \begin{eqnarray}
  h_i -\lambda^2\Delta h_i  = \Phi_0{\hat v}_i\delta(x_0,y_0)\,.   
  \label{L-isotr}
\end{eqnarray}
Here  $\Delta=\nabla^2+\partial^2/\partial z^2$ with 2D Laplacian $\nabla^2$;   ${\hat v}$ is the unit vector along the vortex axis, and $(x_0,y_0)$ are the coordinates in the plane perpendicular to ${\hat {\bm v}}$. For an infinite vortex along $z_0$ in uniform material, the coordinates $(x_0,y_0,z_0)$ are the best, because nothing depends on $z_0$. In the case of a vortex crossing the surface of superconducting half-space, this feature is lost, and the  coordinates $(x ,y ,z )$ with    $z=0$ being  the sample surface are more convenient. The delta-function at the right-hand side (RHS) becomes $\delta(x_0)\delta(y_0)=\delta(x\cos\theta-z\sin\theta)\delta(y)$ where $\theta $ is the angle between $\bm z$ and $\bm v$, the ``tilt" angle, and $y$-axis is chosen to have $v_y=0$. 

The solution of the system (\ref{L-isotr}) of linear  differential equations  is  the sum of its particular solution
and of the solution $\bm h^s$ of the homogeneous equation  with zero RHS. To have a correct singular behavior at the vortex axis, we choose the particular solution as the well-known field of an infinite straight vortex $\bm h^v$:
  \begin{eqnarray}
  \bm h = \bm h^v+\bm h^s\,.    
  \label{hv+hs}
\end{eqnarray}
After taking a  2D $x,y$ Fourier transform of Eq.\,(\ref{L-isotr}) one obtains at $z=0$ (see App. A of Ref.\,\cite{BLK}):
  \begin{eqnarray}
  h^v_x(\bm k) &=& \frac{\Phi_0 \tan\theta}{\lambda^2Q^2}\,,\quad h^v_y=0\,,\quad h_z^v(\bm k) = \frac{\Phi_0 }{\lambda^2Q^2}\,;\label{hxyz}\\
     Q^2&=&\lambda^{-2}+k^2+k_x^2\tan^2\theta\,.    
  \label{Q2}
\end{eqnarray}
%For brevity, we use $\lambda$ as the unit length; in common units all $k$'s in these formulas should be replaced with $\lambda k$.

Further,   the 2D Fourier transform turns the homogeneous Eq.\,(\ref{L-isotr})  into a system of ordinary differential equations for $h_i^s(\bm k,z)$ in the variable $z$,  
  \begin{eqnarray}
  h_i^s\lambda^{-2} +k^2   h_i^s - \partial^2h_i^s/\partial z^2=0\,,   
  \label{linear diff}
\end{eqnarray}
with solutions:
  \begin{eqnarray}
  h^s_i(\bm k,z) = H_i  (\bm k) e^{qz}\,,\qquad q^2=\lambda^{-2}+k^2\,.     
  \label{hs}
\end{eqnarray}
%for the three Fourier components   $h_i^s$. 
Note that  all components of $\bm h^s$ decay exponentially with the characteristic length  $ 1/q = \lambda/\sqrt{1+\lambda^2k^2}$.

Note also that $H_i$ are not independent: by choosing the particular solution as the field of an infinite vortex which obeys div$\,\bm h^v=0$, we impose the same condition on $\bm h^s$: 
  \begin{eqnarray}
  ik_xH_x+ik_yH_y+qH_z=0 \,.    
  \label{divhs}
\end{eqnarray}

The boundary conditions of the field continuity at $z=0$ read in $\bm k$ space:
  \begin{eqnarray}
   ik_x\varphi &=&h_x^v+h_x^s \,,\nonumber\\ 
    ik_y\varphi &=&h_y^v+h_y^s \,, \label{bc}\\ 
    -k \varphi &=&h_z^v+h_z^s \,.\nonumber
    \end{eqnarray}
Along with  Eqs.\,(\ref{hxyz}) and (\ref{divhs}) these conditions give for the external potential 
 \begin{eqnarray}
 \varphi (\bm k)= -\frac{\Phi_0 }{ \lambda^2 k(k+q )(q -ik_x\tan\theta)}\,\label{eq:iso_phi}\,, 
 \end{eqnarray}
and for the coefficients $H_i$:
  \begin{eqnarray}
  H_x &=& - \Phi_0 \frac{(k_y^2+kq)\tan\theta+ik_xq }{ \lambda^2k (k+q)Q^2}  \,,\nonumber\\
 H_y &=& - \Phi_0\frac{ik_y(ik_x\tan\theta  +q) }{ \lambda^2k (k+q)Q^2}  \,, \label{Hi} \\
H_z&=&   \Phi_0\frac{ik_x\tan\theta-k}{ \lambda^2(k +q)Q^2}\,. \nonumber
\end{eqnarray}

%%%%%%%
\subsection{Distribution of the field $\bm {h_z(x,y;z)$}}
%%%%%%

\begin{figure}
\includegraphics[width=9cm]{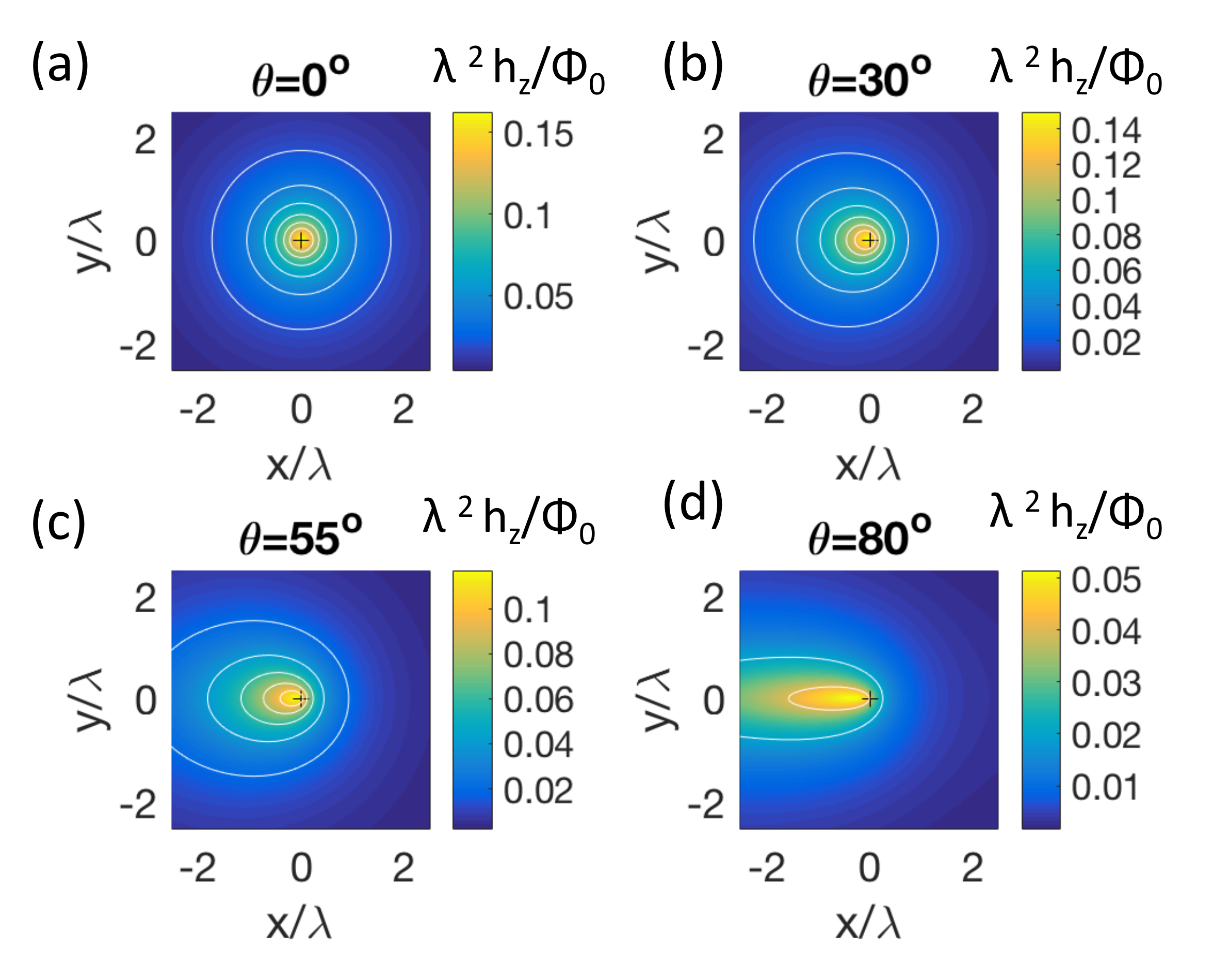}
\caption{ (Color online) Normalized z-component of the magnetic fields $\lambda^2h_z/\Phi_0$ a height $z=0.1\lambda$ above a superconducting  surface for tilted vortices in the isotropic case. 
%The tilt angle of the vortex relative to the sample surface $\theta=0^{\circ}$ (a), 30$^{\circ}$ (b), 55$^{\circ}$ (c), and 80$^{\circ}$ (d). 
The contours of constant $h_z$ (white) are at $\lambda^2h_z/\Phi_0$ = 0.02, 0.04, 0.06, 0.08, 0.10 and 0.12. The `+' symbols mark the centers of the vortex coordinate system where the vortex axis touches the surface.}
\label{fig:isotropic_Hz}
\end{figure}

From the potential (\ref{eq:iso_phi}) we get the $z$ component of the field outside:
  \begin{equation}
 h_z(\bm k,z) =\frac{\Phi_0\, e^{-kz}}{  \lambda^2 (k+q )(q -ik_x\tan\theta)}\,.\label{hz(k,z)} 
 \end{equation}
 In principle, this field can be measured in scanning Hall bar or SQUID experiments.
Figure \ref{fig:isotropic_Hz} shows results of numerical inversion of this Fourier transform to real space.
The vortex fields above the sample surface become  weaker and more elongated in the tilt ($x$) direction as the tilt angle increases.

%%%%%%%
\subsection{Supercurrents    $\bm {J(x,y)}$ at the surface}
%%%%%%%

Supercurrents flowing at the surface affect the order parameter   and the DOS measured by STM. It is not easy to track this connection for arbitrary temperatures. For a qualitative argument we can use the Ginzburg-Landau theory which gives a simple relation for the order parameter suppression by current, $\Delta^2=\Delta_0^2(1-J^2/4J_d^2)$, where $\Delta_0$ corresponds to zero-current and $J_d=c\Phi_0/16\pi^2\lambda^2\xi$ is on the order of the depairing current ($\xi$ is the coherence length) \cite{Abrik}. According to deGennes  the zero-bias density of states $N$ in the vortex vicinity is related to the order parameter as $N(\bm r)/N_0=1-\Delta^2(\bm r)/\Delta_0^2$ \cite{deGennes}. This suggests that the contours $J^2(x,y)$ = const should be close to the DOS contours $N(x,y)$ = const. Of course, the London approach employed here cannot be trusted at distances on the order $\xi$, where the current approaches the depairing value. Still, being interested in a qualitative description of the vortex core shape at the sample surface, one can study the function $J^2(x,y)$. 

% Physically, this is because the quasiparticle energy $\epsilon$ in the presence of superflow  experiences  a shift $\epsilon=\Delta_0-\bm p_F\bm v$ where $\bm p_F$ is the Fermi momentum and $\bm v$ is the superflow velocity \cite{Hess}.
  
  The part of the current at $z=0$ associated with the unperturbed tilted vortex 
has been given in \cite{BLK}: 
 \begin{eqnarray}
 J^v_x(\bm k) &=&\frac{c\Phi_0}{4\pi\lambda^2}\,\frac{i k_y}{ Q^2}\,,\label{eq:jvx}\\
 J^v_y(\bm k ) &=&-\frac{c\Phi_0}{4\pi\lambda^2}\,\frac{i k_x}{Q^2\cos^2\theta }\,.
 \label{eq:jvy} 
\end{eqnarray}
 
The contribution to the current due to the field $\bm h^s$  of Eq.\,(\ref{hs}) at $z=0$   follows from Maxwell equations:
 \begin{eqnarray}
\frac{4\pi } {c }J^s_x(\bm k)  &=&  ik_yH_z(\bm k )-qH_y(\bm k )\,.
 \label{eq:jx} \\
\frac{4\pi } {c } J^s_y(\bm k ) &=& qH_x(\bm k )- ik_xH_z(\bm k ) \,,\label{eq:jy}
\end{eqnarray}
Hence, we can evaluate the current value  at the surface in real space:
  \begin{equation}
  J^2(x,y)=(J_x^s+J_x^v)^2+(J_y^s+J_y^v)^2.
  \label{eq:jsq}
  \end{equation}  
  
\begin{figure}
\includegraphics[width=8.5cm]{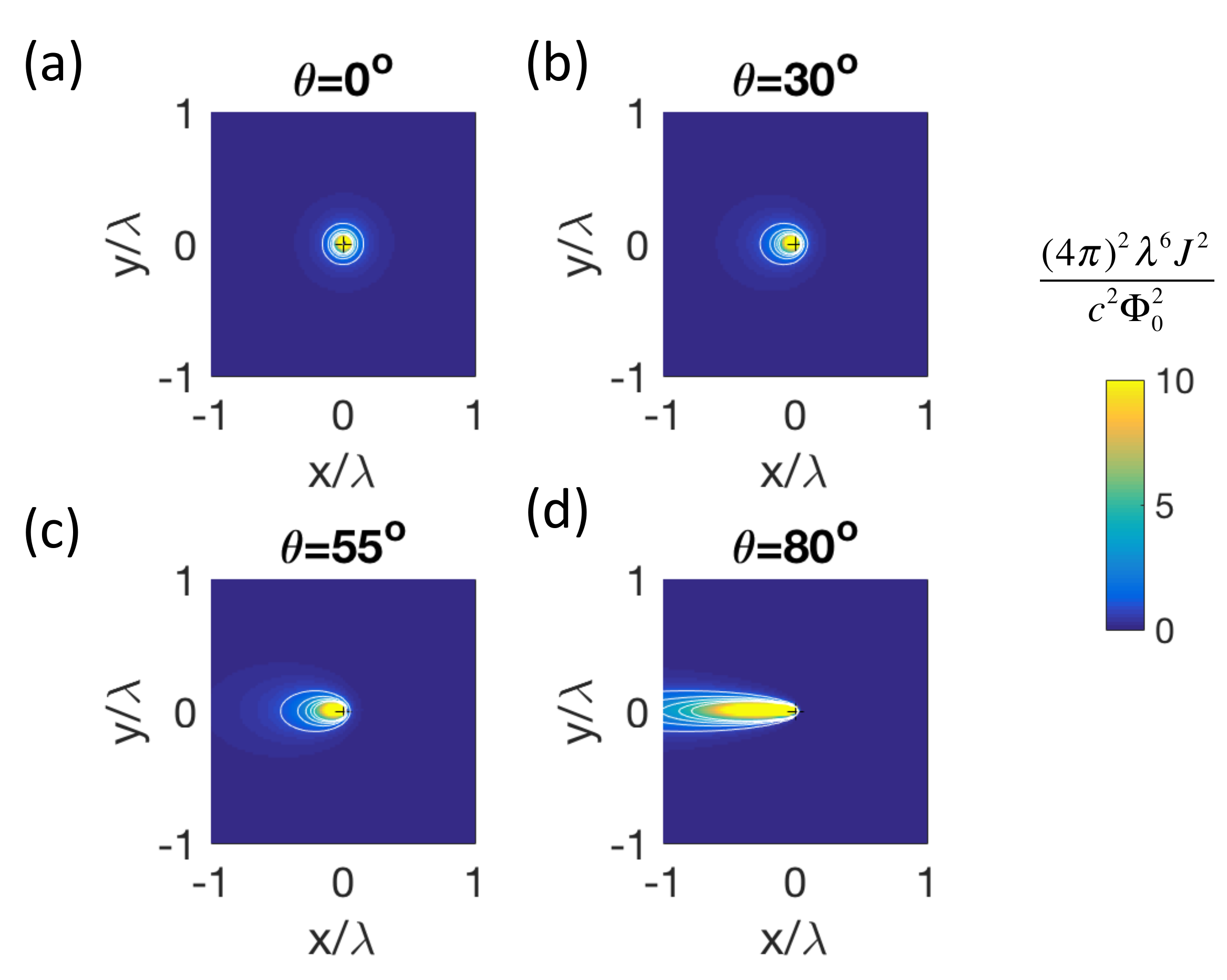}
\caption{(Color online) Normalized absolute value squared of the vortex currents $(4\pi)^2\lambda^6J^2/c^2\Phi_0^2$ at the superconducting surface in the isotropic case. The tilt angle of the vortex relative to the sample surface $\theta=0^{\circ}$ (a), 30$^{\circ}$ (b), 55$^{\circ}$ (c), and 80$^{\circ}$ (d). The contours of constant $J^2$ (white) are at $(4\pi)^2\lambda^6J^2/c^2\Phi_0^2$ = 1, 2, 3, 4, and 5. The `+' symbol marks the center of the vortex coordinate system where the vortex axis touches the surface. }
\label{fig:isotropic_jsq}
\end{figure}

Some numerical evaluations of Eq.\,\ref{eq:jsq} are displayed in Figure \ref{fig:isotropic_jsq}.  For these calculations we applied a high frequency filter by multiplying the right hand sides of Eqs.\, (\ref{eq:jvx})--(\ref{eq:jy}) by $e^{-k^2 dz^2}$ with $dz=0.01\lambda$. This damps out high frequency artifacts at $x=0$ and $y=0$ without significantly effecting the low frequency properties of the solutions. The false color scale in Fig.\,\ref{eq:jsq} is saturated at $(4\pi)^2\lambda^6J^2/c^2\Phi_0^2$=10. Since physically the core shape (as observed in e.g. STM) is determined by a contour where $J$ reaches the depairing value, the contours $J^2(x,y)$=constant will also give the contours of DOS$(x,y)$=constant: the observed vortex cores will become more elongated along the tilt ($x$) direction as the tilt angle increases. To avoid misunderstanding, we stress that the white curves in Fig.\,\ref{fig:isotropic_jsq} are contours $J^2(x,y)$=const, not the stream lines of the vector $\bm J(x,y)$.

%%%%%% 
   \section{Uniaxial crystal with  surface at  $\bm{ab}$ plane}
%%%%%%%

The general case of an anisotropic half-space superconductor with an  arbitrary plane surface and arbitrarily oriented vortex has been considered in Ref.\,\cite{KSL}.
 Here, we are interested in the   surface coinciding with the $ab$ plane, Section III.A of  \cite{KSL}. In this case, the frame $x,y,z$ coincides with crystal's $a,b,c$, and the mass tensor is diagonal: $m_{xx}=m_{yy}=m_a$, $m_{zz}=m_c$, the ``effective masses" are normalized  $m_a^2m_c=1$, and the anisotropy parameter $\gamma=\sqrt{m_c / m_a}=\lambda_c/\lambda_{ab}$. In what follows the unit of length is given by $\lambda=(\lambda_{ab}^2\lambda_c)^{1/3}$.

The basic scheme of the solution is the same as in the isotropic case: one has to solve the anisotropic London equations \cite{K81} inside and to match them to solutions of the Maxwell equations for the field outside. Without going into formal details (for which   readers are referred to Ref.\,\cite{KSL}) we note a relevant point: while solving the  system of  London equations for the surface contribution to the internal field in the form  
$h^s_i(\bm k,z) = H_i  (\bm k) e^{qz}$ we obtain a  system of linear homogeneous equations for $H_i  (\bm k)$, the determinant of which must be zero. This  gives possible values of the parameter $q$. After straightforward algebra one obtains two positive roots: 
  \begin{eqnarray}
 q_1 =\sqrt{ \lambda_{ab}^{-2}+ k^2} \,,\qquad  q_2 =\sqrt{\lambda_{ab}^{-2}+\gamma^2k^2}  \,  . \label{q12}  
 \end{eqnarray}
Hence, instead of one mode of the field decay of the  isotropic case, we have now two such modes.  
The pre-factors $\bm H^{(1)}$ and $\bm H^{(2)}$  are given by:
  \begin{eqnarray}
&&H^{(1)}_x  =  H^{(1)}_y\frac{k_x}{k_y}  =  H^{(1)}_z\frac{ik_xq_1}{ k^2}\,, \label{H1}\\
&&H^{(1)}_z =\Phi_0\frac{ ik_x\tan\theta -k}{( k+q_1) d_1},\label{H1z} \\
&&d_1=1+\lambda_{ab}^2(k^2+k_x^2\tan^2\theta); \label{d1}\\
&&H^{(2)}_x  = - H^{(2)}_y\frac{k_y}{k_x} = -\Phi_0\frac{k_y^2\tan\theta}{k^2d_2} \,,\quad H^{(2)}_z=0,\label{H2}\\
&& d_2=1+\lambda_c^2 k^2+\lambda_{ab}^2 k_x^2\tan^2\theta \label{d2}\,.  
 \end{eqnarray}
 
 The boundary conditions of the field continuity at $z=0$ now read:
   \begin{eqnarray}
   ik_x\varphi &=&h_x^v+H_x^{(1)}+ H_x^{(2)}\,,\label{bc-new-x}\\ 
    ik_y\varphi &=&h_y^v+H_y^{(1)}+ H_y^{(2)} \,, \label{bc-new-y}\\ 
    -k \,\varphi &=&h_z^v+H_z^{(1)}  \,.\label{bc-new-z}
    \end{eqnarray}
%Since $H_x^{(1)}$ and $H_y^{(1)}$ are expressed in terms of $H_z^{(1)}$
 
 The 2D Fourier components of the  field $\bm h^v$ at $z=0$ are given in App. A of Ref.\,\cite{BLK}:
   \begin{eqnarray}
&&h^{v}_x  =  \Phi_0\tan \theta[1+\lambda_{ab}^2(k_x^2\tan^2\theta+k_y^2)+\lambda_{c}^2 k_x^2]/d \,,\qquad\label{hvx}\\
&&h^{v}_y  =  \Phi_0\tan \theta (\lambda_c^2 -\lambda_{ab}^2)k_xk_y /d \,, \label{hvy} \\
%&&d_1=1+\lambda_{ab}^2(k^2+k_x^2\tan^2\theta); \label{d1}\\
&&h^{v}_z  =  \Phi_0/[1+  \lambda_{ab}^2(k_x^2\sec^2\theta+k_y^2)] \,,\quad d=d_1d_2\,.\label{hvz} 
%&& d =[1+\lambda_{ab}^2(k_x^2\sec^2\theta+k_y^2)](1+\lambda_{c}^2 k^2+\lambda_{ab}^2k_x^2\tan^2\theta) \,.\nonumber\\
%\label{hv}  
 \end{eqnarray}

The condition div$\,\bm h^s=0$ at $z=0$ translates to $ik_x(H_x^{(1)}+ H_x^{(2)})+ik_y(H_y^{(1)}+ H_y^{(2)})+q_1H_z^{(1)} =0$, so that one easily excludes all $\bm H$'s from the system (\ref{bc-new-x})-(\ref{bc-new-z}) to obtain:
  \begin{eqnarray}
 \varphi &=& -\frac{\Phi_0\,e^{-kz}}{\lambda_{ab}^2 k(k+q_1)(q_1-ik_x\tan\theta)}\,\label{phi}\,.
%  q_1^2&=&\frac{1+m_ak^2}{m_a}. \label{q1}
\end{eqnarray}
Note that $\lambda_c$ does not enter this expression. Hence, the outside field depends only on $\lambda_{ab}$.
It is worth noting that if one replaces the vortex as the field source with some external source, the response field outside  also does not depend on $\lambda_c$  \cite{Meissner}.

From the potential we get:
  \begin{equation}
 h_z = -k\varphi(\bm k)=\frac{\Phi_0\, e^{-kz}}{\lambda_{ab}^2 (k+q_1)(q_1-ik_x\tan\theta)}\,.\label{hz} 
 \end{equation}\\
For $k\to 0$, $q_1=1/\lambda_{ab}$   so that the total flux $h_z(k=0)=\Phi_0$, as it should. 
%In common units we have:
%  \begin{eqnarray}
% h_z &=& \frac{\Phi_0\, e^{-kz}}{(\lambda_{ab}k+p_1)(p_1-i\lambda_{ab}k_x\tan\theta)}\,,\label{hz_comm}\\
% p_1^2&=& 1+\lambda_{ab}^2k^2 . \label{p1}  
% \end{eqnarray}

%**********
\begin{figure}
\includegraphics[width=8.5cm]{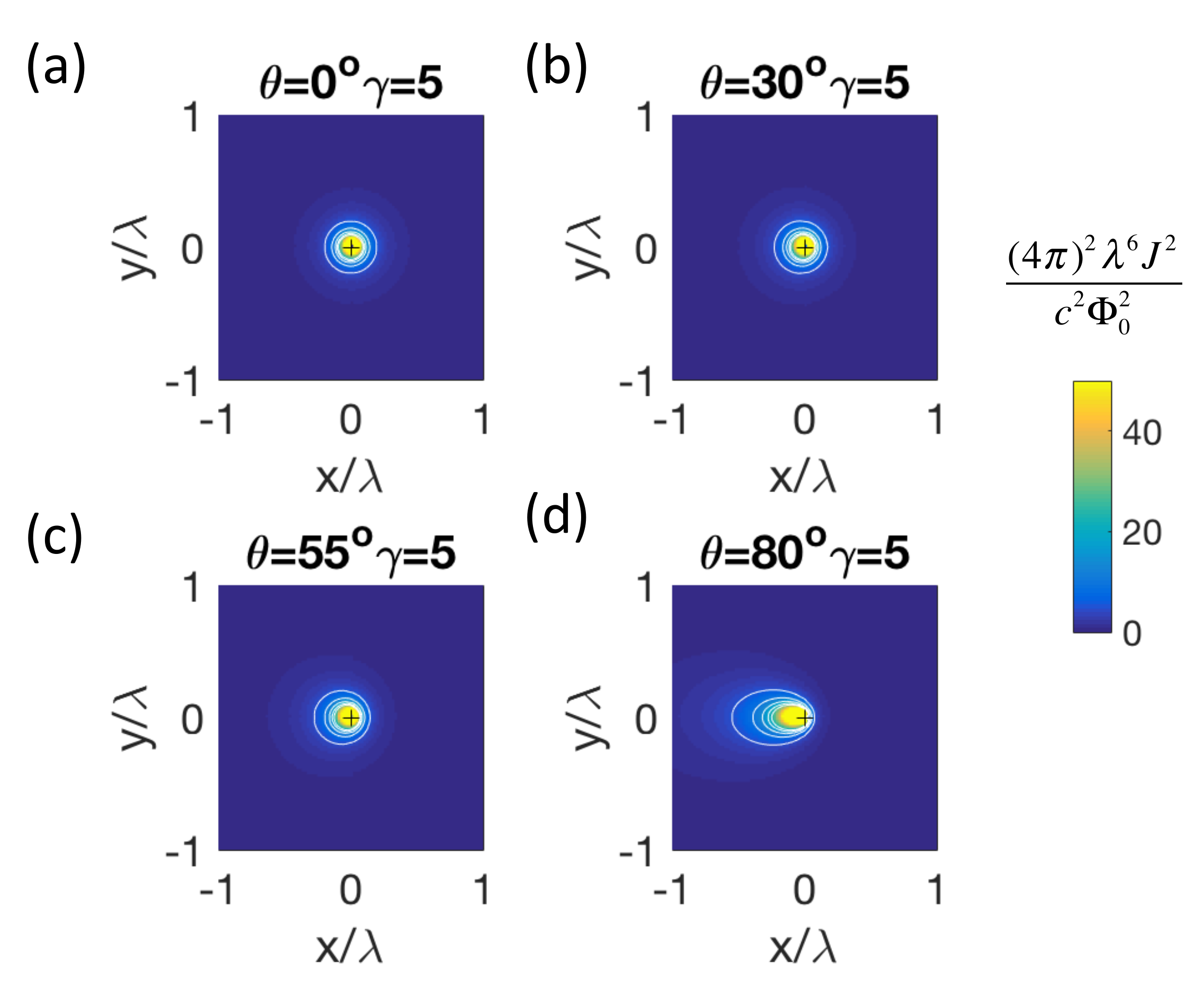}
\caption{(Color online) Normalized absolute value squared of the vortex currents $(4\pi)^2\lambda^6J^2/c^2\Phi_0^2$ at the superconducting surface in the uniaxial anisotropic case with $\gamma=5$. The tilt angle of the vortex relative to the sample surface $\theta=0^{\circ}$ (a), 30$^{\circ}$ (b), 55$^{\circ}$ (c), and 80$^{\circ}$ (d). The false colormap is saturated at $(4\pi)^2\lambda^6J^2/c^2\Phi_0^2=50$. The contours of constant $J^2$ (white) are at $(4\pi)^2\lambda^6J^2/c^2\Phi_0^2$ = 5, 10, 15, 20, and 25. The `+' symbol marks the center of the vortex coordinate system. }
\label{fig:anisotropic_jsq}
\end{figure}

\subsection{Surface currents}

 As before, the current consists of the vortex and surface contributions, ${\bm J}^v$ and ${\bm J}^s$. 
%Currents can be written as  sum of vortex and surface contributions: 
      The surface contribution is given by
 \begin{eqnarray}
\frac{4\pi } {c } J^s_x(\bm k)  &=& ik_yH_z^{(1)} -q_1H_y^{(1)}  - q_2H_y^{(2)} \,,
 \label{jsx} \\
\frac{4\pi } {c } J^s_y(\bm k ) &=&q_1H_x^{(1)}+ q_2H_x^{(2)}   - ik_xH_z^{(1)} \,.
 \label{jsy}
\end{eqnarray}
%The factor $4\pi/c$ is omitted.

For a tilted vortex, the currents at the surface are given in Appendix A of Ref.\,\cite{BLK}:
 \begin{eqnarray}
\frac{4\pi } {c }J^v_x &=& \Phi_0\, \frac{ik_y [1+\lambda^2_c(k^2+k_x^2\tan^2\theta)] }{d_1\,d_2}\,,\label{jvx}\\
\frac{4\pi } {c } J^v_y &=& -\Phi_0\, \frac{ik_x  [1+ \lambda_{ab}^2(\sin^2\theta+\gamma^2\cos^2\theta) (k^2+k_x^2\tan^2\theta)]  }{ d_1\,d_2 \cos^2\theta}. \nonumber\\
 \label{jvy}
\end{eqnarray}
 
   One can now evaluate numerically $J^2(x,y)=(J_x^s+J_x^v)^2+(J_y^s+J_y^v)^2$   at the surface.   
 Results are shown in Figure \ref{fig:anisotropic_jsq}. We again applied a high frequency filter $e^{-k^2 dz^2}$ with $dz=0.01\lambda$ to damp out high frequency oscillations at $x=0$ and $y=0$. The false color scale in Fig. \ref{fig:anisotropic_jsq} is saturated at $(4\pi)^2\lambda^6J^2/c^2\Phi_0^2$=50. Note that the current densities are higher and the elongation of the vortex core along the tilt axis at high tilt angles is less pronounced as compared with the isotropic case (Fig. \ref{fig:isotropic_jsq}).

\begin{figure}
\includegraphics[width=8.5cm]{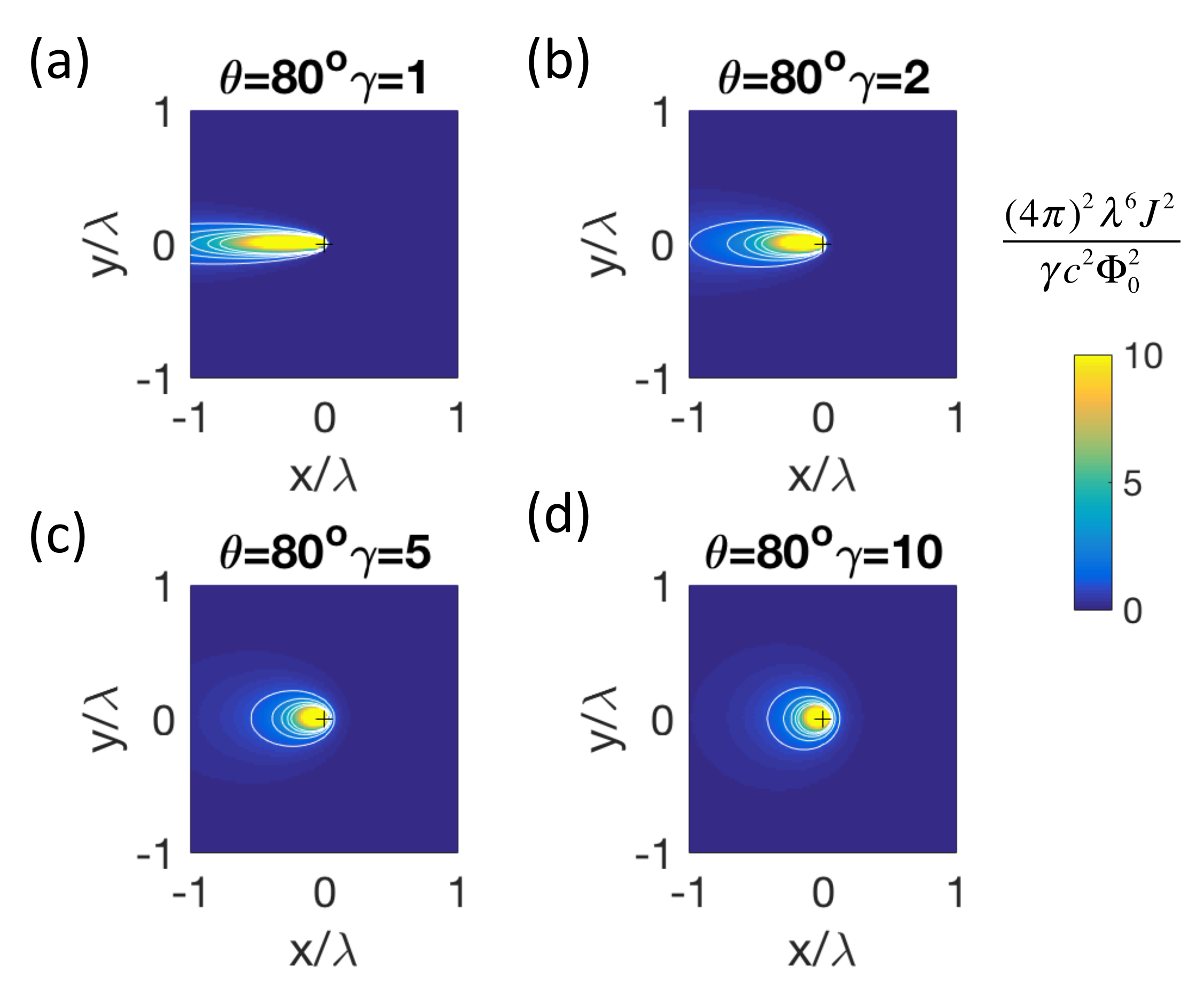}
\caption{(Color online) Normalized absolute value squared of the vortex currents $(4\pi)^2\lambda^6J^2/\gamma c^2\Phi_0^2$ at the superconducting surface in the uniaxial anisotropic case for a tilt angle of 80$^{\circ}$. The anisotropy parameter is  $\gamma$= 1 (a), 2 (b), 5 (c), and 10 (d). The false color scales are saturated at $(4\pi)^2\lambda^6J^2/\gamma c^2\Phi_0^2$= 10. The contours of constant $J^2$ (white) are at $(4\pi)^2\lambda^6J^2/\gamma c^2\Phi_0^2$ = 1, 2, 3, 4, 5. The `+' symbol marks the center of the vortex coordinate system. }
\label{fig:anisotropic_vs_gamma}
\end{figure}

The systematic behavior of the vortex core with uniaxial anisotropy is illustrated in Figure \ref{fig:anisotropic_vs_gamma}.

%  We expect these contours be symmetric relative to $y\to -y$, but asymmetric relative to $x\to -x$.

%Doing this calculation one can drop $4\pi/c$ and $\Phi_0$.

\section{Discussion}

Numerical analysis of the expressions given above show that the external magnetic fields from vortices become weaker as the tilt angle increases, at the same time as the vortex shape becomes more elongated in the tilt direction (Fig. \ref{fig:isotropic_Hz}). In contrast, the peak absolute values of the surface supercurrents are relatively insensitive to tilt angle, while the vortex core elongation increases with tilt angle (Fig. \ref{fig:isotropic_jsq}).  For a uniaxial superconductor the surface currents become stronger with higher anisotropy, but the vortex cores become less elongated (Fig.'s \ref{fig:anisotropic_jsq}, \ref{fig:anisotropic_vs_gamma}). This, at first sight, is surprising but could be understood qualitatively by comparing tilted vortices near the surface in the isotropic and anisotropic cases. Since there the currents must be parallel to the surface,  in isotropic materials the surface causes a strong distortion of the currents in its vicinity as compared to the bulk. On the other hand, in an anisotropic uniaxial sample with the $ab$ surface, the unperturbed bulk current planes are already tilted toward $ab$ due to anisotropy, so that the distortion caused by the surface is getting weaker with increasing anisotropy. In the limit $\gamma\gg 1$, the surface distortion disappears altogether, which we in fact see in our simulations.

Experimental tests of these effects would be a challenge with existing trilayer \cite{kirtley2016rsi} or Dayem bridge \cite{veauvy2002rsi} SQUID microscopes, which have spatial resolution of somewhat less than 1$\mu$m, while superconducting penetration depths are typically 0.1$\mu$m. However, recent SQUID-on-a-tip sensors \cite{vasyukov2013nnano} may have the spatial resolution required. Of course, STM easily has the spatial resolution to look for the vortex elongations predicted here.

%Questions about STM: sometimes we seem to see tilted vortex cores (ovals, comet-tails) in other situation the tilted vortices bend to exit normal to the surface.

\section{Acknowledgements}

The authors thank H. Suderow for many helpful discussions.  V.K. was supported by the U.S. Department of Energy, Office of Science, Basic Energy Sciences, Materials
Sciences and Engineering Division. The Ames Laboratory is
operated for the U.S. DOE by Iowa State University under
Contract No. DE-AC02-07CH11358.  J.K. was supported by Stanford University.

\end{document}